\documentclass[aip,jcp,preprint]{revtex4-1}
\usepackage{graphicx,color}
\usepackage{dcolumn}% 
\usepackage{bm}%%
\usepackage{epstopdf}
\usepackage{color}

\draft % marks overfull lines with a black rule on the right

\begin{document}

% Use the \preprint command to place your local institutional report
% number in the upper righthand corner of the title page in preprint mode.
% Multiple \preprint commands are allowed.
% Use the 'preprintnumbers' class option to override journal defaults
% to display numbers if necessary
%\preprint{}

%Title of paper
\title{Atomistic molecular dynamics simulations of shock compressed quartz}

% repeat the \author .. \affiliation  etc. as needed
% \email, \thanks, \homepage, \altaffiliation all apply to the current
% author. Explanatory text should go in the []'s, actual e-mail
% address or url should go in the {}'s for \email and \homepage.
% Please use the appropriate macro foreach each type of information

% \affiliation command applies to all authors since the last
% \affiliation command. The \affiliation command should follow the
% other information
% \affiliation can be followed by \email, \homepage, \thanks as well.
\author{M~R~Farrow}
%\email[]{Your e-mail address}
%\homepage[]{Your web page}
%\thanks{}
%\altaffiliation{}
\affiliation{Department of Physics, University of York, Heslington, York, YO10
5DD, United Kingdom}

\author{M~I~J~Probert}
\affiliation{Department of Physics, University of York, Heslington, York, YO10
5DD, United Kingdom}

\date{\today{}}

\begin{abstract}
Atomistic non-equilibrium molecular dynamics (NEMD) simulations of shock wave compression of quartz have been performed using the so-called BKS semi-empirical potential of van Beest, Kramer and van Santen\cite{KramerFVV91} to construct the Hugoniot of quartz. Our scheme mimics the real world experimental set up by using a flyer-plate impactor to initiate the shock wave and is the first shock wave simulation that uses a geometry optimised system of a polar slab in a 3-dimensional system employing periodic boundary conditions. Our scheme also includes the relaxation of the surface dipole in the polar quartz slab which is an essential pre-requisite to a stable simulation. The original BKS potential is unsuited to shock wave calculations and so we propose a simple modification. With this modification, we find that our calculated Hugoniot is in good agreement with experimental shock wave data up to 25 GPa, but significantly diverges beyond this point. We conclude that our modified BKS potential is suitable for quartz under representative pressure conditions of the Earth core, but unsuitable for high-pressure shock wave simulations.  We also find that the BKS potential incorrectly prefers the $\beta$-quartz phase over the  $\alpha$-quartz phase at zero-temperature,  and that there is a $\beta \to \alpha$ phase-transition at 6 GPa. 
\end{abstract}

% insert suggested PACS numbers in braces on next line
% insert suggested keywords - APS authors don't need to do this
%\maketitle must follow title, authors, abstract, \pacs, and \keywords
\maketitle
% References should be done using the \cite, \ref, and \label commands
\section{Introduction}
Silicates are abundant in the Earth's interior where they experience high-pressures of up to 136 GPa, therefore their behaviour at high-pressure is of great interest to researchers in many areas including physics, astrophysics and geosciences. A well known silicate, quartz, or silicon dioxide ($\mathrm{SiO_2}$) is found in the $\alpha$-quartz phase at standard temperature and pressure with each silicon atom being 4-fold coordinated with the oxygen atoms and has a crystal structure made up of $\mathrm{SiO_4}$ tetrahedra building blocks. Quartz has many high-pressure, high-temperature polymorphs;  $\beta$-quartz (a high-temperature polymorph), coesite (a high-pressure polymorph) and stishovite (a high-temperature, high pressure polymorph). Many of these polymorphs are thermodynamically close in energy to each other. 

Unsurprisingly, quartz and its polymorphs have been subject to extensive theoretical and experimental studies \cite{Wackerle62,HemleyJMMM88,MurashovS98} though there is still some discussion as to the phase changes that quartz undergoes with pressure, such as the recent work by Atkins and Ahrens who used early shock wave data along with recent discoveries of post-stishovite phases of $\mathrm{SiO_2}$ to reassign the phase transition regions along the Hugoniot of quartz \cite{AkinsA02}. However, theoretical shock wave studies have been limited \cite{BarmesSM06,PhysRevLett.67.3559}.  Static pressure, diamond anvil cell experimental X-ray diffraction data for quartz have been gathered by Hemley \emph{et al}  \cite{HemleyJMMM88}. They found that their samples showed the onset of amorphisation between 25 to 35 GPa (at 300K). This amorphisation has also been shown in molecular dynamics (MD) simulations \cite{PhysRevLett.67.3559}. 

The simulation technique and choice of empirical potential is essential, and for silicates, many researchers choose the well known semi-empirical potential of van Beest, Kramer, and van Santen; the so-called BKS potential \cite{KramerFVV91}. This potential is predominantly used for equilibrium simulations \cite{MurashovS98,KimizukaKK03}, and includes point charges and a Buckingham-type pair potential which becomes unphysical (infinitely attractive) at small interatomic separations. This is  catastrophic for the high-pressures achieved during shock wave simulations, and so researchers have corrected for this unphysical response, such as  Barmes \emph{et al} \cite{BarmesSM06} who have used a second-order polynomial fit, whereas Guissani and Guillot \cite{GuissaniG96} have added a Lennard-Jones type potential.

 There are a number of alternative simulation methods available to obtain the Hugoniot; Brennan and Rice \cite{Brennan:2003fk} have adapted the methodology of Erpenbeck \cite{PhysRevA.46.6406} in which a point on the Hugoniot curve is calculated from single simulations (MD or Monte-Carlo) that each calculate an equation of state point. Maillet \emph{et al} have created the so-called Uniaxial Hugoniostat method \cite{MailletMSRLGH01} which is an equilibrium molecular dynamics method that uses perturbed equations of motion that obey the Rankine-Hugoniot relations. In this way, the equilibrium MD simulations result in the long-time relaxed structure after shock compression. A modified Hugoniostat has been used with some success by Ravelo \emph{et al} and Barmes \emph{et al} \cite{RaveloHGL04,BarmesSM06}.

In this study we have developed atomistic non-equilibrium molecular dynamics (NEMD) techniques to simulate the non-equilibrium state experienced during shock compression for materials that contain point charges such as in the BKS potential. We use atomistic NEMD simulations to give direct information on the mechanisms at play at the atomic scale, and the underlying mechanisms of the phase transformations. We believe that this is the first complete shock wave simulation study that uses a geometry optimised system of a polar slab along with an impactor to generate the shock wave in a 3-dimensional system employing periodic boundary conditions (PBC).  Within the 3-dimensional PBC the computationally efficient Ewald summation scheme was used to correctly handle the long-range Coulomb interactions. We demonstrate our simulation technique by computing the Hugoniot of quartz up to pressures of 400 GPa.  Such pressures are extremely high, and do not correspond to equilibrium Earth mantle conditions at which quartz is naturally occurring, but they are in the range found for meteorite impacts \cite{MeloshC05}. 

The paper is structured as follows: First, we discuss the details of the simulations; the initial optimisation of the atomic structure to form a semi-infinite slab model, followed by the subsequent relaxation of the dipole moment of the polar surface.  We then apply the scheme to perform shock wave simulations on quartz, and discuss the results and implications for our choice of semi-empirical potential.

\section{Simulation details} 
All the simulations were performed using molecular dynamics in the micro-canonical ensemble (NVE) with periodic boundary conditions (PBC) and used a Velocity-Verlet integrator.   The interatomic potential used was the so called BKS potential of van Beest, Kramer and van Santen \cite{KramerFVV91}: 
\begin{widetext}
\begin{equation}
\label{bks}
U_{bks}(r) = \sum_{i>j}\frac{q_{\alpha _i}q_{\beta _j}}{r_{\alpha i \beta j}} - \sum_{i>j}\left [A_{\alpha _i \beta _j}exp(-b_{\alpha _i \beta _j}r_{\alpha _i \beta _j}) - \frac{C_{\alpha _i \beta _j}}{r_{\alpha _i \beta _j}^6}\right ]
\end{equation}
\end{widetext}
where $\alpha$ and $\beta$ are atomic species and the first term describes the long-ranged electrostatic interaction between atoms $i$ and $j$ which is determined by the species-dependent effective charges $q_{\alpha}$ and $q_\beta$. The second term is the short-ranged interactions in a Buckingham-type form, where $A_{\alpha \beta}$, $b_{\alpha \beta}$, and $C_{\alpha \beta}$ are constants derived from fitting to Hartree-Fock \emph{ab-initio} calculations on aluminophosphates and selected empirical measurements. These force-field parameters have been shown to be reasonably successful in describing the dynamic and structural properties of quartz and some of its polymorphs \cite{MurashovS98,KimizukaKK03}.  The parameters used by van Beest \emph{et al} are reproduced in table \ref{bks_params}.  A  cut-off radius of  $6.0$ \AA\  was used for both the silicon-oxygen bonds and oxygen-oxygen bonds, respectively.  The problem of long-range Coulomb forces with PBCs was handled by using an Ewald summation. 

\begin{figure}
   \includegraphics[scale=0.75]{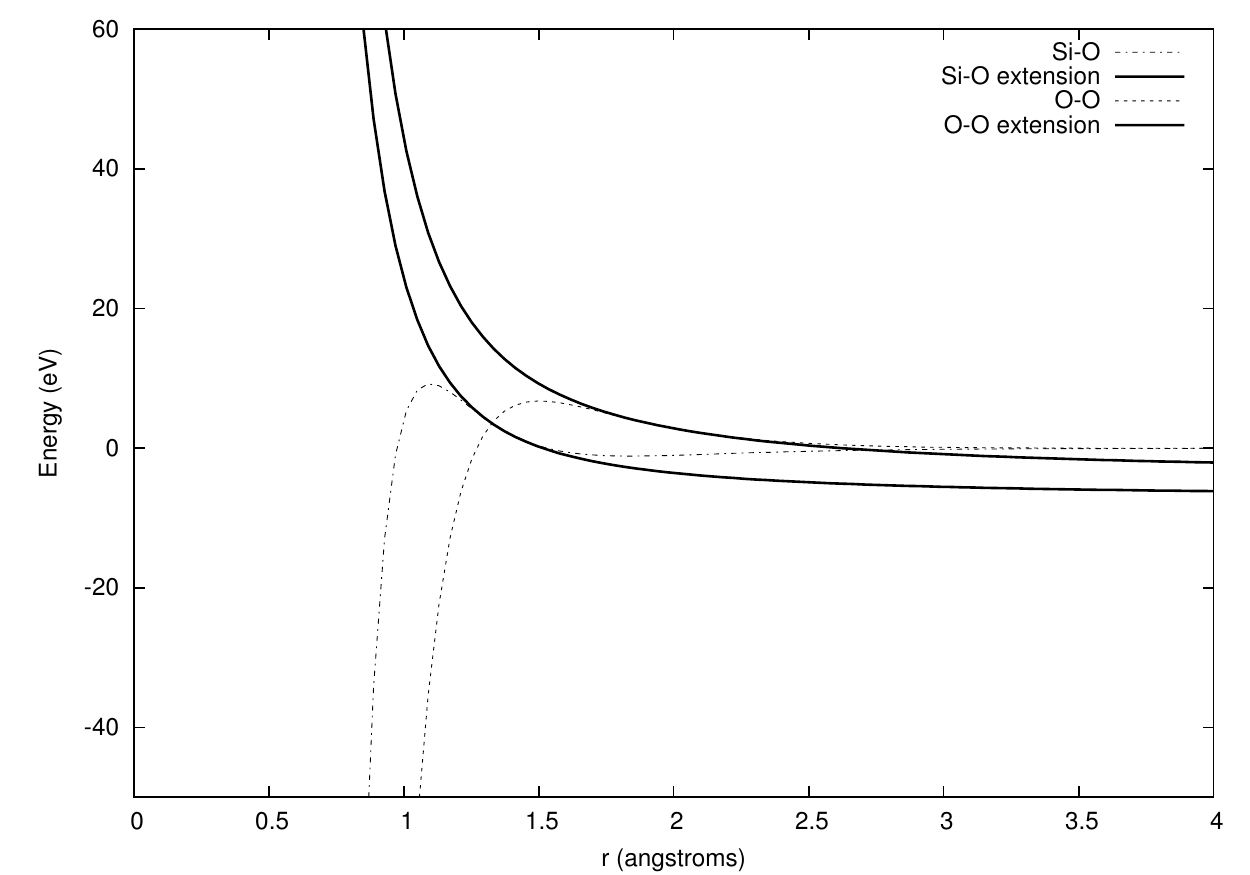} 
     \caption{Plot of the interatomic potential used in this work. Dashed lines show the unphysical behaviour of the original BKS potential at small interatomic distances.}
     \label{bks_ext_fig}
  \end{figure}

 \begin{table}%[H] add [H] placement to break table across pages
   \caption{Reproduction of the fit parameters of van Beest \emph{et al} for the BKS potential \cite{KramerFVV91}}
 \label{bks_params} 
 \begin{ruledtabular}
 \begin{tabular}{lcccc}
  $\alpha_i\beta_j$ & $A_{\alpha_i \beta_j}$(eV) & $b_{\alpha_i \beta_j}($\AA$^{-1})$ & $C_{\alpha_i \beta_j}(eV $\AA$^6)$  & $q$ \\
    \hline
     Si-O & 18003.7572 & 4.87318 & 133.5381 & $q_{Si} = 2.40$\\
     O-O &1388.77300 & 2.76000& 175.0000 & $q_{O}=-1.2$ \\
 \end{tabular}
 \end{ruledtabular}
 \end{table}
The second term in the BKS potential has an unphysical maximum at small bond lengths.  The large compressions that can occur during shock wave simulations cause the BKS potential to fail (become infinitely attractive), so it is wise to correct for this unphysical response.  We have replaced this part of the BKS potential at the point of inflection with the following polynomial form:
\begin{equation}
\label{bks_ext}
U_{ext}(r) =\sum_{i>j} \frac{D_{\alpha _i\beta _j}}{r_{\alpha _i \beta _j}^2} + \frac{E_{\alpha _i\beta _j}}{r_{\alpha _i \beta _j}^{6}} + F_{\alpha _i\beta _j}
\end{equation}
where $D_{\alpha \beta}$, $E_{\alpha \beta}$ and $F_{\alpha \beta}$ are calculated analytically to match the BKS pair potential and its first and second derivatives, at the point of inflection. Table \ref{bks_ext_table} gives numerical values of these parameters.  This $2-6$ form was chosen as it provided a Lennard-Jones like repulsion at short distances whilst having a numerically convenient form and matches smoothly onto the original BKS Buckingham term.  Figure \ref{bks_ext_fig} shows the form of the Buckingham term in BKS potential and shows the smooth transition region from BKS to our extension.  Therefore, our modified short-ranged interaction potential has the following form: 
\begin{equation}
U(r) = \left \{
\begin{array}{ll} 
 U_{ext}(r) & r \leq r_{inf}\\
U_{bks}(r) & r > r_{inf} \leq r_{cut} \\ 
 0 & r > r_{cut}  
\end{array} \right .
\end{equation} 
where $r_{inf}$ is the point of inflection.  For the Si-O and O-O parts of the BKS potential, $r_{inf}$ was $1.35$\AA\ and $1.98$\AA, respectively.
 
\begin{table}
   \caption{Numerical values of the parameters used for correcting the BKS pair potential at small bond lengths}
  \label{bks_ext_table}
  \begin{ruledtabular} 
   \begin{tabular}{cccc} 
    $\alpha_i \beta_j$ & $D_{\alpha  \beta}(eV $\AA$^2)$ & $E_{\alpha \beta}(eV $\AA$^6)$ & $F_{\alpha \beta}(eV)$\\
   \cline{1-4} \cline{1-4}
     Si-O & 24.1700 & 23.8086 & -3.5872 \\
     O-O & 12.3435 & 18.9662 & -6.9426 \\
   \end{tabular}
   \end{ruledtabular}
  \end{table}
    
\begin{figure}
  \includegraphics[scale=1.5]{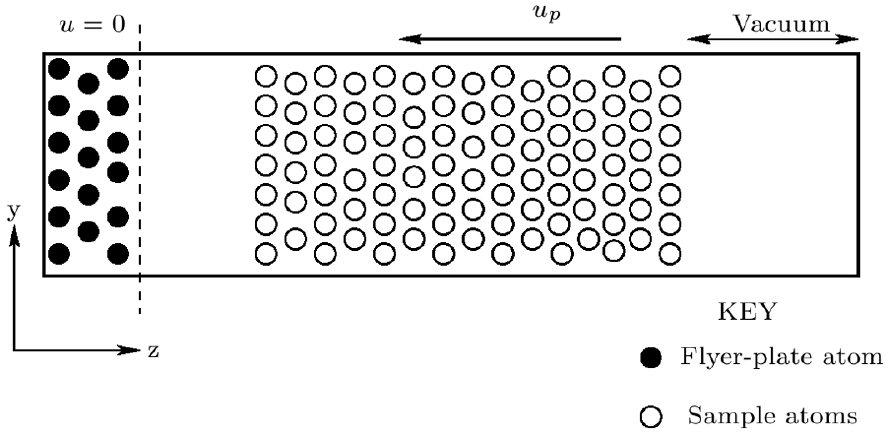} 
  \caption{Schematic of a shock wave system. Sample atoms are given an initial velocity, $u_p$, towards the stationary flyer-plate. Periodic boundary conditions are used, therefore a vacuum region removes unwanted interactions between the end surfaces.}
  \label{schematic}
\end{figure}

The systems studied contained 1584 and 3600 atoms of $\alpha$-quartz ($4\times4\times11$ and $4\times4\times25$ unit cells, respectively).  The systems were equilibrated to 300K using a Berendsen weak-coupling thermostat then further equilibrated for 2 ps using standard NVE dynamics before the shock wave was initiated.  Initially we generated a shock wave in the system by giving all of the atoms in the system a ``piston velocity" of $-u_p$ towards a so-called momentum mirror \cite{Holian88}.  This created a shock wave that propagated in the positive z-direction at velocity $u_s$.  The momentum mirror technique essentially places a perfectly reflecting surface at the origin, that reverses the position and velocity of an atom that passes  through it during a molecular dynamics time-step.  Our momentum mirror was located at the z=0 plane.  However, this technique was found to yield compressions that were unphysical and resulted in extreme deformations of the system.  It also resulted in predominantly using our extension to the BKS potential and so this was the first evidence that the BKS potential was unsuitable for shock wave simulations.  We replaced the infinitely stiff surface of the momentum mirror with a small block of quartz, very much like the flyer-plate used in shock wave experiments; for this reason it is referred to as the flyer-plate.  The flyer-plate interacts with the incoming atoms through the same empirical pair potential and therefore was much ``softer" than the momentum mirror. By Galilean invariance, this simulation set up mimics the incoming flyer-plate impactor in real world experiments. In order for the flyer-plate to remain intact, and to create a shock wave in the more numerous ``sample" atoms (i.e. those atoms in the system that were not the flyer-plate) each atom in the flyer-plate was given a large mass of 100 times its usual mass.   The piston velocities were in multiples of the sound velocity, $c_0$, and the simulations ran until the system reached the shocked state.  A schematic of the system can be seen in figure \ref{schematic}.  As can be seen in the schematic, a vacuum gap  between the end of the sample atoms and the start of the flyer-plate atoms is essential to ensure that no interactions via PBC affected the simulations.  We set our gap to be larger than the cut-off radius of the BKS potential (6 \AA), however, long-range Coulomb forces were still present which created a large dipole moment in this polar system. 

\begin{figure}
   \includegraphics[scale=0.5]{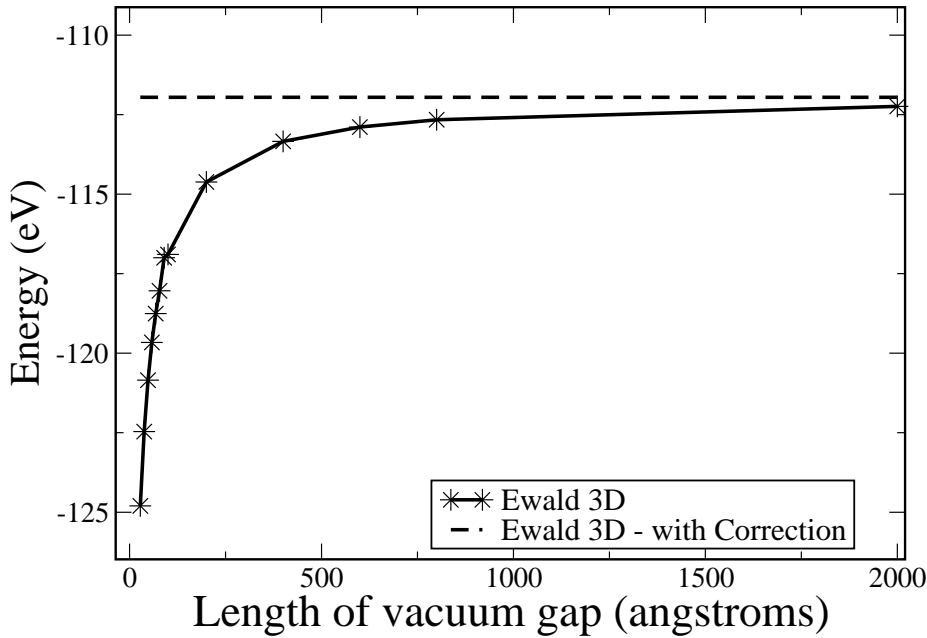} 
      \caption{The energy of Ewald 3D with correction (EW3DC) converges to the aperiodic long-range limit of the 3D Ewald summation technique.}
\label{ewald_fig}
\end{figure}

 This was accounted for by performing a geometry optimisation on the system and correcting the Ewald summation to that of the aperiodic 3D limit.  This correction, proposed by Yeh and Berkowitz \cite{YehB99}, is more computationally efficient than using a 2D Ewald summation alone. The method, denoted here as EW3DC simply applied as an additive correction to the 3D Ewald summation energy and forces; the energy correction, J(\textbf{M},P) is dependent on the total dipole moment  \textbf{M} and the shape (P). Our system had the geometry of a rectangular plate (P=R) and hence our energy correction term is given by:
\begin{equation}
\label{correction_term}
J(\mathbf{M},R) = \frac{2 \pi}{V}M_z^2
\end{equation}
and \textbf{M} is given by:
\begin{equation}
\label{dipole_moment}
\mathbf{M} = \sum_{i=1}^N q_i\mathbf{r}_i
\end{equation}
A correction is also applied to the force calculation, obtained by differentiation of the energy term.  Figure \ref{ewald_fig} shows the convergence to the aperiodic long-range limit of the 3D Ewald summation using EW3DC.  It was clear that a small vacuum gap (but greater than the cut-off of the Buckingham term of the BKS potential) would be acceptable.  

\subsection{Geometry optimisation and static compression} 
\begin{figure}
   \includegraphics[scale=0.5]{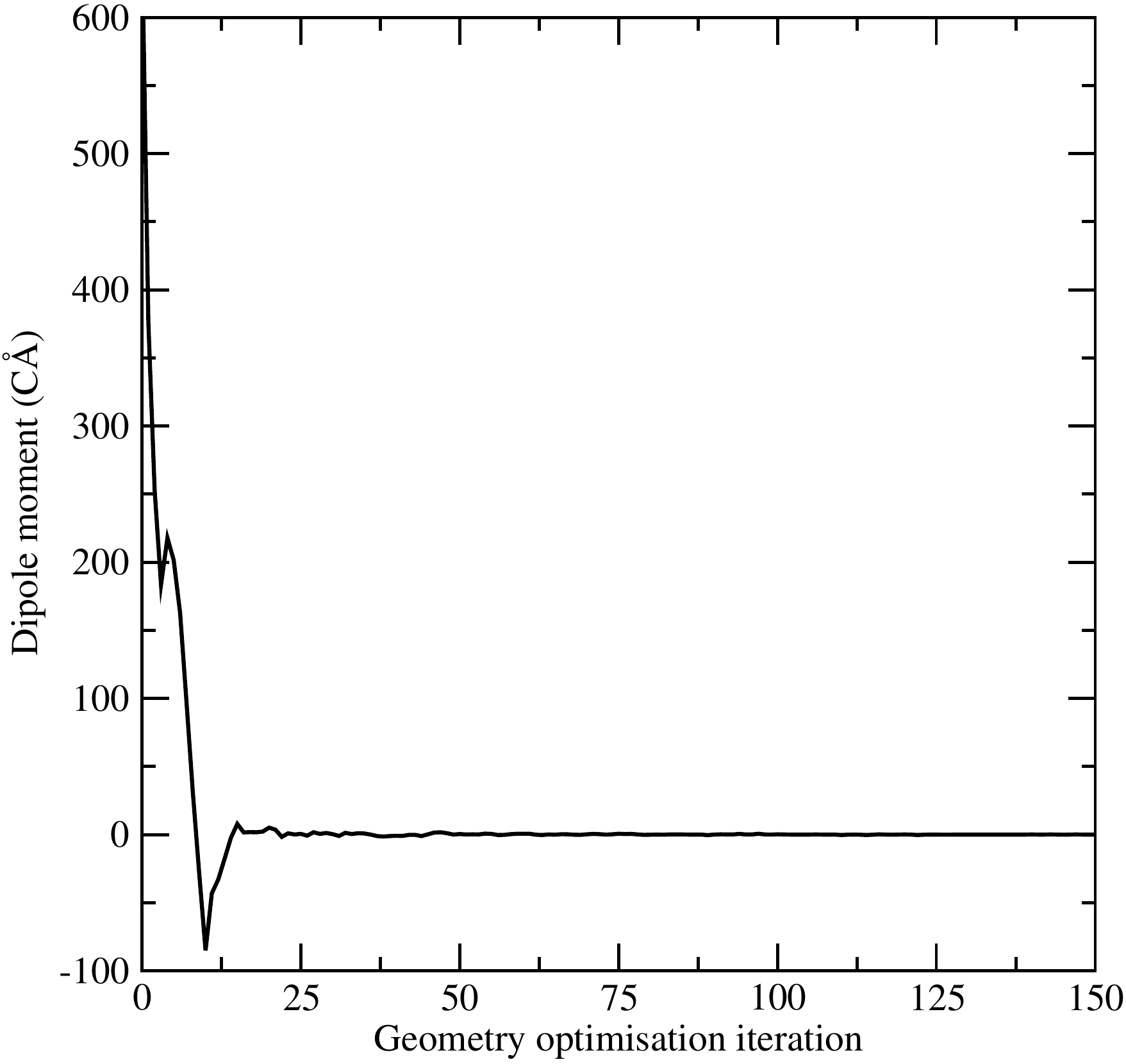} 
   \caption{Dipole moment with geometry optimisation steps. The dipole moment of the system is negligible after 50 steps.}
   \label{Dipole}
 \end{figure}

Before starting any MD simulations, it is essential to have a relaxed and stable system. Making a quartz slab by cleaving bulk quartz creates a large and unphysical dipole moment, which strongly effects the energetics of the system. Hence the systems were first relaxed using the BFGS algorithm along with the EW3DC which removed the dipole moment that was initially present.  Figure \ref{Dipole} shows how the optimisation removed the dipole moment of the system after 50 steps and from there continued to relax the system to its final state. The final relaxed structure had negligible dipole moment across the slab.

\begin{figure}[htbp]
   \centering
   \includegraphics[scale=0.45]{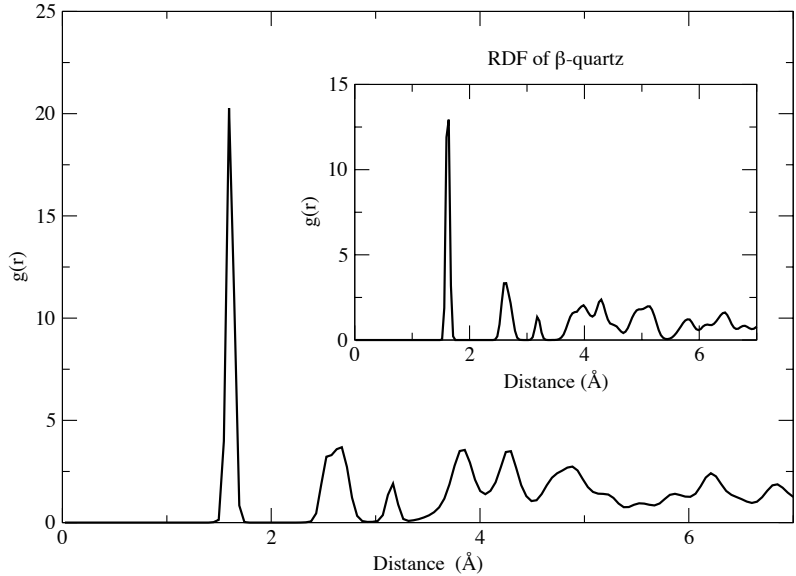} 
   \caption{Radial Distribution Function of quartz after geometry optimisation. As can be seen from the RDF of $\beta$-quartz (inset) the system is in the $\beta$-quartz phase prior to shock compression}
   \label{rdf_quartz}
 \end{figure}

\begin{figure}
   \includegraphics[scale=0.5]{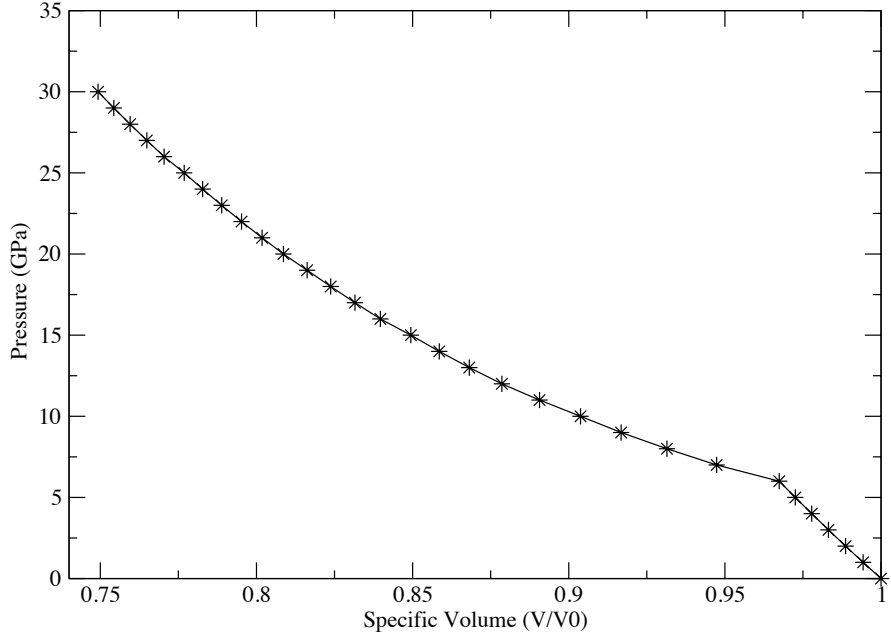} 
   \caption{Pressure--Specific-volume plot for static compression of quartz. Structure is initially in the $\beta$-quartz phase and transforms to $\alpha$-quartz at 6GPa.}
   \label{Static_compression}
 \end{figure}

\begin{figure}
   \includegraphics[scale=0.5]{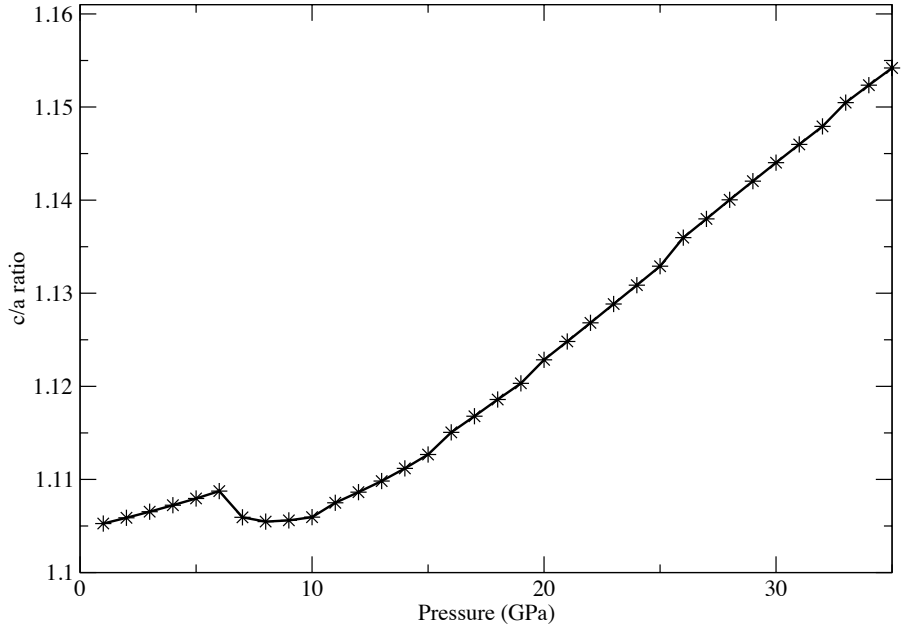} 
   \caption{c/a ratio during static compression. The discontinuity at 6GPa corresponds to the phase transformation from $\beta \to \alpha$ quartz.}
   \label{c_over_a}
 \end{figure}

Figure \ref{rdf_quartz} shows the radial distribution function (RDF) of the relaxed system prior to shock wave simulation initiation. The structure that was found was that of $\beta$-quartz, a high-temperature polymorph of quartz. To investigate why this polymorph was observed a series of static pressure calculations were performed, isotropically compressing bulk quartz.  Figure \ref{Static_compression} shows the pressure-volume curve of the static compression simulations and figure \ref{c_over_a} shows the $c/a$ ratio of the relaxed structure as a function of pressure. It can be seen that at 6 GPa the system undergoes a phase change.  Analysis of the structure of the system prior to this phase change showed that the low pressure phase is $\beta$-quartz and the high pressure phase is $\alpha$-quartz. van Beest \emph{et al}\cite{KramerFVV91} also noted that their BKS potential gave $\beta$-quartz as the zero pressure low temperature phase. In this regard, it is worth noting that there are other potentials available.  Using the so-called TTAM potential \cite{TsuneyukiTAM88} that has a similar form to BKS (although a little more computationally expensive) researchers have found that the temperature of the $\alpha$-$\beta$ phase transition is reproduced accurately \cite{herzbach:124711}.  The Tangey and Scandolo (TS) potential \cite{tangney:8898} has also been shown to give good agreement with the experimental c/a ratio and the $\alpha$-$\beta$ phase transition in quartz. All of the potentials have their good and bad points, and interested readers are referred to the study by Herzbach \emph{et al} \cite{herzbach:124711} for a comparison between the popular silicate potentials. 

\begin{figure}
    \includegraphics[scale=1.3]{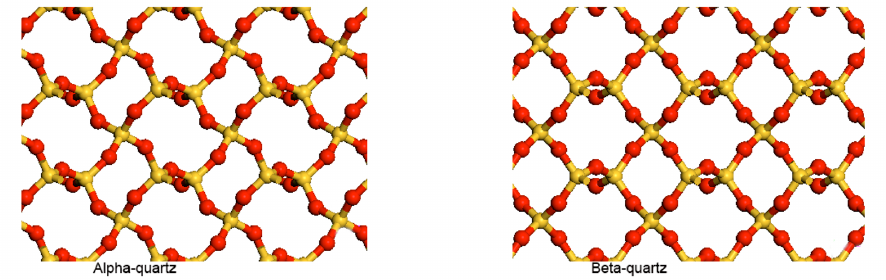} 
   \caption{Quartz structures: left, $\alpha$-quartz and right, $\beta$-quartz. The higher symmetry of $\beta$-quartz can be seen. }
   \label{quartz_structures}
 \end{figure}

The two structures of quartz are shown in figure \ref{quartz_structures}.  This unphysical transition means that at low temperatures a pressure of 6 GPa or greater is required to maintain $\alpha$-quartz in this phase when using the BKS potential.  We saw no evidence for the onset of amorphisation between 20 and 30 GPa which has been reported previously \cite{PhysRevLett.67.3559,PhysRevB.53.107}.
 
\section{Results} 
\begin{figure}
   \includegraphics[scale=0.5]{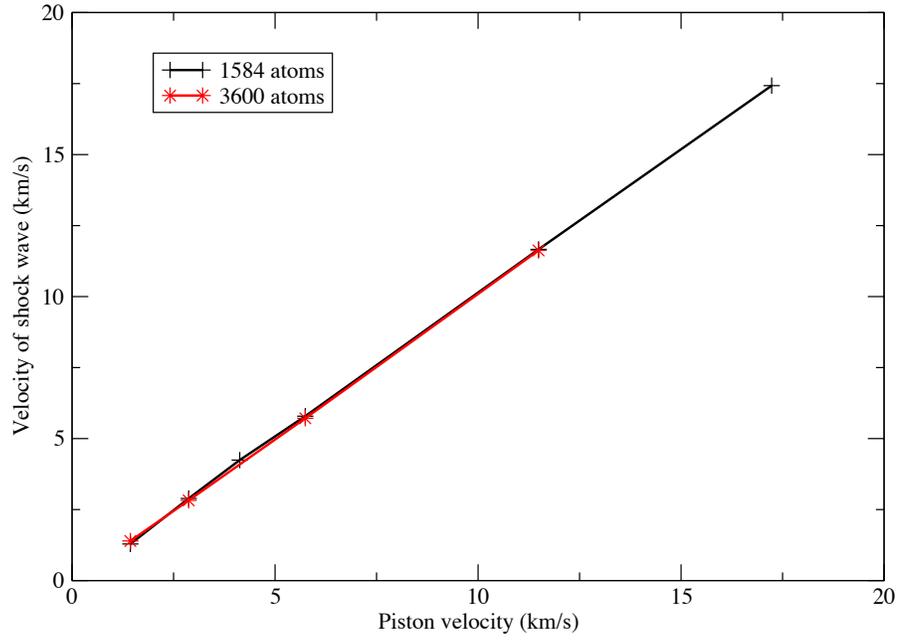} 
   \caption{$u_p$-$u_s$ Hugoniot of quartz for 1584 and 3600 sample atom systems. A linear dependence of shock wave velocity with piston velocity is clear, with no evidence for any finite-size effects.}
   \label{UsUp_quartz}
\end{figure}

\begin{figure}
   \includegraphics[scale=0.5]{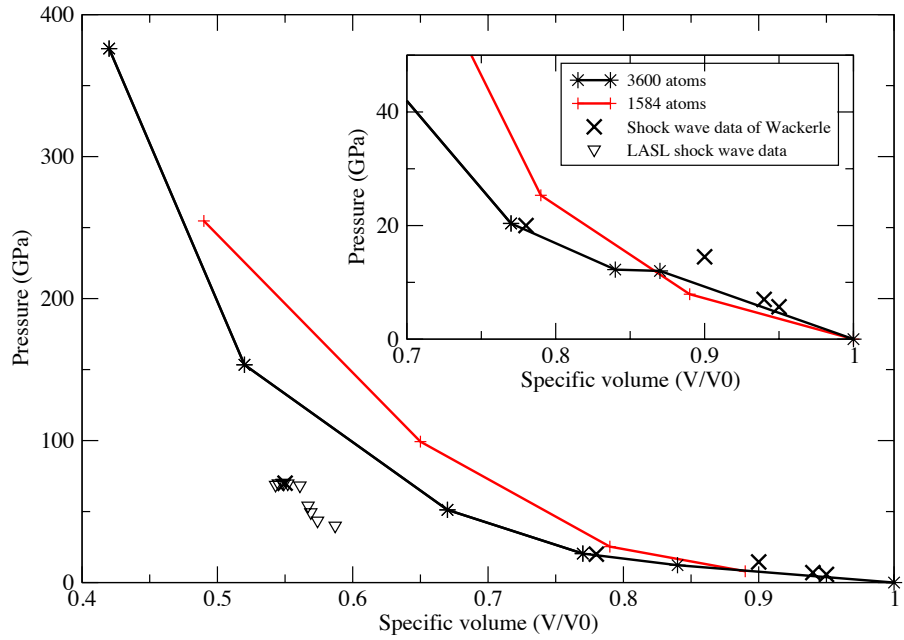} 
   \caption{P-V Hugoniot of $\alpha$-quartz. Experimental data from Wackerle \cite{Wackerle62} and Marsh. \cite{Marsh}}
   \label{hug_quartz}
\end{figure}

\begin{figure}
   \includegraphics[scale=0.40]{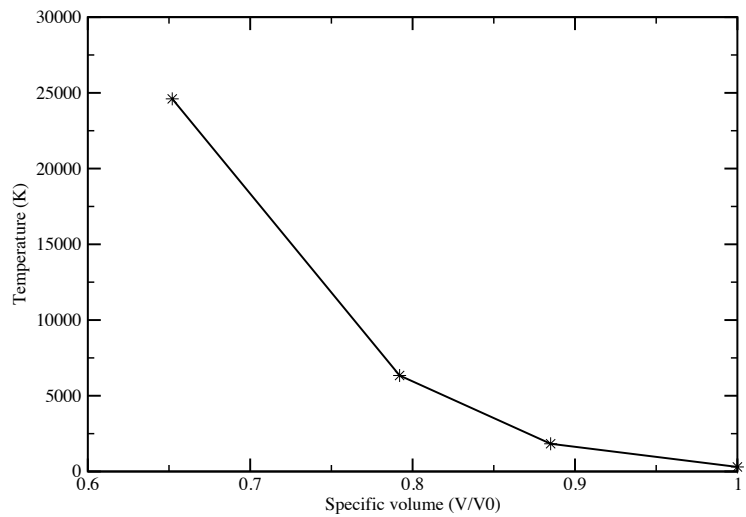} 
   \caption{Temperature variation along the Hugoniot of quartz (1584 atoms) }
   \label{Temp_hug_quartz}
\end{figure}

Figure \ref{UsUp_quartz} shows the piston ($u_p$) and shock ($u_s$) velocity Hugoniot.  As can be seen from the figure, the piston and corresponding shock velocities are large but are easily achieved in the laboratory. Such shock conditions also occur naturally, for instance, at the site of a meteorite impact --- which is how the high pressure polymorph stishovite was first discovered. Meteorites have a mean impact velocity \cite{MeloshC05} of between 17 -20 km/s but can be as low as 10 km/s --- hence this is an interesting regime to explore by simulation.  A linear dependance of piston ($u_p$) to shock ($u_s$) velocity can clearly be seen. The calculations are repeated for both 1584 and 3600 atom systems and there is no evidence of finite size effects.
The corresponding pressure-volume Hugoniot for the shock wave compression calculations are shown along with experimental data from Wackerle \cite{Wackerle62}  and LASL data of Marsh \cite{Marsh} in figure \ref{hug_quartz}.  Our Hugoniot gives reasonable agreement to the experimental data at low pressures (small piston velocities) up to 25 GPa, but at higher pressures it deviates significantly. Whilst there is some evidence of a finite-size effect in this figure, it is unlikely that this is sufficient to explain the discrepancy at high pressures. Instead, this deviation is most likely due to the high-pressure polynomial fit to the BKS potential; as at these pressures this part of the pair-potential term is sampled frequently and thus we can conclude that the fit yields a response that is too stiff. As this polynomial fit does not have any physical justification, it could be possible to ``tune" the fitting parameters to yield a much better agreement with the experimental Hugoniot.  However, this would effectively be making a high-pressure potential due to the frequency of it being sampled during the simulations and the form may not be transferrable to other silicates or polymorphs of quartz. Therefore, it is safer to conclude that at such high-pressures, that are far from equilibrium, the BKS potential breaks down and should not be considered a useful potential for these conditions.  It is possible that a re-parameterisation of the potential at these high-pressure conditions incorporating this sort of empirical shock-wave data may yield a better potential for shock wave simulations \cite{Farrow:2009}.  As it stands, our results appear to be reliable up to 25 GPa, which is sufficient for modelling the terrestrial geothermal states of quartz, but progressively deteriorate at higher pressures. The rapid increase in temperature along the Hugoniot of quartz (1584 atoms) can be seen in figure \ref{Temp_hug_quartz}.  At such high temperatures it is unsurprising that the resultant structure is amorphous as the temperature is vastly greater than the melting point of any of the quartz polymorphs.

\begin{figure}
   \includegraphics[scale=0.45]{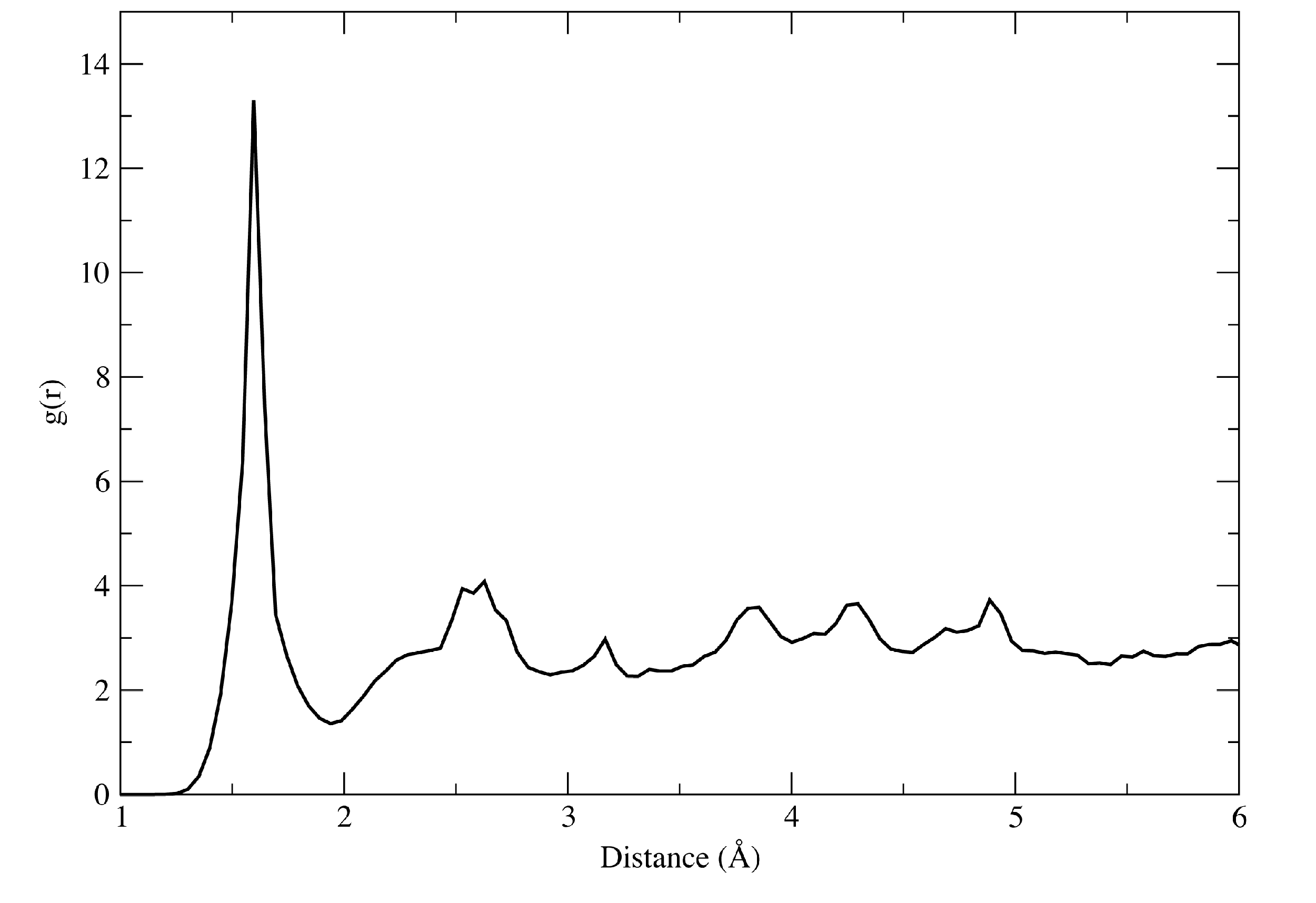} 
   \caption{Radial Distribution Function of $\alpha$-quartz at maximum shock compression using a 3 km/s piston velocity. The amorphous structure can be clearly seen.}
   \label{Shocked_rdf}
\end{figure}

At maximum shock compression the system was in an amorphous state as can be seen from the RDF in figure \ref{Shocked_rdf}.  This was generated from a shock wave that was initiated by a 3 km/s piston velocity (corresponding to half the speed of sound in $\alpha$-quartz).  This shocked structure does not correspond to stishovite, although it is possible that if it was allowed to equilibrate over a long period of time with slow cooling, that stishovite might form. However, such long time scales are not achievable in this type of molecular dynamics. We can however derive some knowledge about the adiabatic response of the system after the shocked state has been generated by calculating the decompression isentrope. 

\begin{figure}
   \includegraphics[scale=0.4]{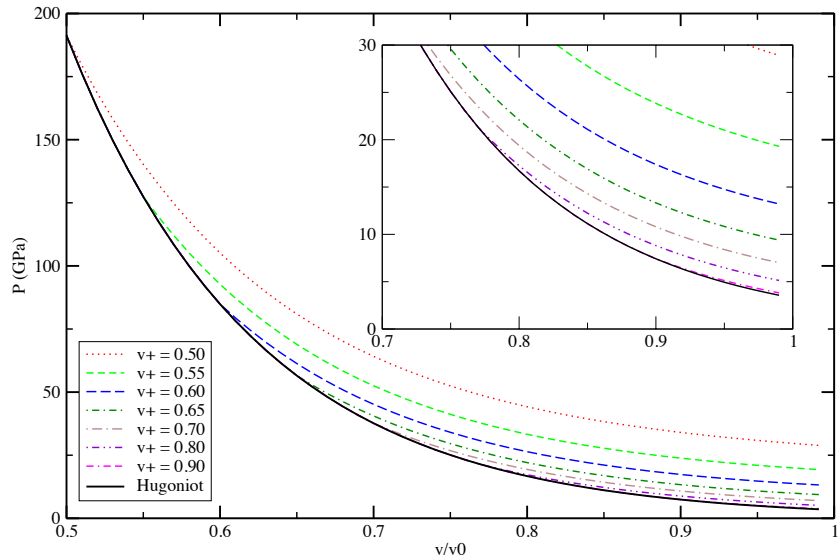} 
   \caption{Fit to 3600 atom quartz Hugoniot along with calculated isentropes from various specific volume at release points, v+. Inset: The decompression response over the geophysically relevant region for quartz.}
   \label{isentropes}
\end{figure}

The calculated  Hugoniot is well represented by the equation: 
\begin{equation}
P = \mathbf{A}exp\left ( \mathbf{B}v \right )
\end{equation}
where $\mathbf{A}=11112$ GPa, $\mathbf{B}=-8.1245$ and $v$ is the specific volume (V/V0). This form was used to calculate the release isentropes from various release points, $\mathrm{v}+$, using the following relation between the Hugoniot and an isentrope \cite{Davison}:
\begin{equation}
P^\gamma(v) =  \chi_c(v)\left [ p^+ + \int_{v^+}^v \frac{\kappa_c(v')}{\chi_c(v')}dv' \right ]
\end{equation}
where
\begin{equation}
\kappa_c(v) = \frac{1}{2}\frac{\gamma_r}{v_r}P^H(v) + \left [ 1 - \frac{1}{2}\frac{\gamma_r}{v_r}(v_r-v) \right ] \frac{dP^H(v)}{dv}
\end{equation}
and
\begin{equation}
\chi_c(v) = exp\left [ \frac{\gamma_r}{v_r}(v^+ - v)\right ]
\end{equation}
$P^\gamma(v)$ is the isentropic pressure, $P^H(v)$ is the Hugoniot pressure, $p^+$ is the Hugoniot pressure at specific volume, v+ at release, centered on specific volume $v_r$. $\gamma(v)$ is the Gruneisen parameter; taken to be $0.659$ from Quareni \emph{et al} \cite{QuareniM89}.   Figure \ref{isentropes} shows our calculated isentropes from the fit to the Hugoniot.  There is little deviation from the Hugoniot at small compressions, up to $v+=0.80$, corresponding to an initial Hugoniot pressure of 20 GPa. However at larger compressions (and higher pressures) the deviation grows considerably. 

\clearpage 
\section{Conclusion} 
We have performed atomistic non-equilibrium molecular dynamics simulations of shock wave compression on quartz. We chose the widely used BKS interatomic potential. However in order to avoid complications with the unphysical behaviour at high pressures in the pair-interaction part of the BKS potential, we created a simple ($2,6$) polynomial expression with an analytically determined fit to the BKS potential at the point of inflection and its derivatives. Our systems were periodic in 3-dimensions but had a vacuum gap between the end of the sample atoms and the start of the flyer-plate atoms as a means of preventing unwanted interactions due to periodic boundary conditions.  We used the Ewald summation correction for 3D systems as proposed by Yeh and Berkowitz as a means of allowing a geometry optimisation to be successfully performed, in order to eliminate the dipole moment caused by cleaving a slab from bulk quartz.  The optimised structure was found to be that of the high-temperature polymorph $\beta$-quartz and not $\alpha$-quartz. We found a phase change from $\beta \to \alpha$ quartz at 6 GPa by performing static compression calculations.   Our analysis of the radial distribution functions showed that the shock compression of quartz formed an amorphous phase. 
 
The calculated Hugoniot of quartz gave a reasonable agreement to experimental data for modest shock velocities, up to a pressure of 25 GPa. For the terrestrial geotherm, this corresponds to a depth of approximately 700 km which is beyond the region in which quartz is believed to be found. The agreement  significantly deviated at higher pressures and larger shock velocities. This was reasoned as due to a breakdown of the BKS potential and to the polynomial fit (for high-pressures) that gave a response that was too stiff.  Using an alternative potential (such as TTAM or TS) may avoid this problem, however they  should first be investigated for their suitability. A re-parameterisation of the BKS  potential for high-pressures would be expected to provide a suitable potential for shock wave simulations. 

We conclude that the BKS potential, in the modified form we propose, is suitable for quartz under representative pressure conditions of the Earth core, but is unsuitable for high-pressure shock wave simulations. 

% If you have acknowledgments, this puts in the proper section head.
\begin{acknowledgments}
% put your acknowledgments here.
The authors would like to thank the Engineering and Physical Sciences Research Council (EPSRC) for financial support. 

\end{acknowledgments}

\end{document}